\documentclass[superscriptaddress,amsmath,A4,pra,twocolumn,showpacs]{revtex4} 
\usepackage{amssymb}
\usepackage{amsthm}
\usepackage{graphicx}
\usepackage{color}
\usepackage{young}

\input{qcircuit}

\newcommand{\tcb}[1]{\textcolor{black}{#1}}

\newcommand{\bra}[1]{\langle#1|}
\newcommand{\ket}[1]{|#1\rangle}

\newcommand{\ketbra}[2]{{\ket{#1}\bra{#2}}}
\newcommand{\Bra}[1]{\langle \! \langle#1|}
\newcommand{\Ket}[1]{|#1\rangle \! \rangle}
\newcommand{\BraKet}[2]{\langle \! \langle #1|#2 \rangle \! \rangle}
\newcommand{\KetBra}[2]{{\Ket{#1}\Bra{#2}}}

\newcommand{\hilb}[1]{\mathcal{#1}}

\def\qed{$\,\blacksquare$\par}
\def\<{\langle}\def\>{\rangle}
\DeclareMathOperator{\Tr}{Tr}

\DeclareMathOperator{\spn}{span}

\newcommand{\chu}{\mathcal{U}}

\newtheorem{theorem}{Theorem}

\begin{document}

\title{Probabilistic storage and retrieval of qubit phase gates}

\author{Michal Sedl\'ak}
\affiliation{RCQI, Institute of Physics, Slovak Academy of Sciences, D\'ubravsk\'a cesta 9, 84511 Bratislava, Slovakia}
\affiliation{Centre of Excellence IT4Innovations, Faculty of Information Technology, Brno University of Technology, Bo\v zet\v echova 2/1, 612 00 Brno, Czech Republic }

\author{M\'ario Ziman}
\affiliation{RCQI, Institute of Physics, Slovak Academy of Sciences, D\'ubravsk\'a cesta 9, 84511 Bratislava, Slovakia}
\affiliation{Faculty of Informatics,~Masaryk University,~Botanick\'a 68a,~60200 Brno,~Czech Republic}

\begin{abstract}
Probabilistic storage and retrieval (PSR) of unitary quantum dynamics is possible with exponentially small failure probability with respect to the number of systems used as a quantum memory [PRL 122, 170502 (2019)]. Here we study improvements due to a priori knowledge about the unitary transformation to be stored. In particular, we study $N \rightarrow 1$ PSR of qubit phase gates, i.e. qubit rotations a round $Z$ axis with an unknown angle, and show that if we access the gate only $N$-times, the optimal probability of perfect retrieving of its single use is $N/(N+1)$. We propose a quantum circuit realization for the optimal protocol and show that programmable phase gate [PRL 88, 047905 (2002)] can be turned into $(2^k-1)\rightarrow 1$  optimal PSR of phase gates and requires only $k$ CNOT gates, while having exponentially small failure probability in $k$.
\end{abstract}

\maketitle

\section{Introduction}
Discovery of Shor's algorithm \cite{shor} boosted research investigating capabilities of quantum systems for computation and information processing. In analogy with classical computers people envisioned a quantum computer, which would have at its heart a quantum processor realizing a fixed unitary transformation on data and program quantum bits (qubits). Ideally the transformation should be universal, i.e. by choice of the state of the program register the machine could be programmed to perform any desired (unitary) transformation on the data qubits. However, Nielsen and Chuang \cite{nielsen1} proved that perfect (error free) implementation of $k$ 
linearly independent unitary transformations requires at least $k$ dimensional program register. This observation has lead to formulation of the no-programming theorem.

Consequently, we either look for deterministic, but approximate processors \cite{dariano2005}, or we design probabilistic processors \cite{processor}, which perform the desired operation exactly and signalize it, but they often have to signalize a failure as well. Although some upper bounds on the achievable performance of approximate or probabilistic processors exist \cite{kubicki} there are still gaps between them and performance of the processors that were found so far \cite{multimeter,multimeter1,garcia2006,portishizaka0,portIshizaka1,portStrelchuk1,portstrelchuk2}.

The task of storage and retrieval of unitary transformations addresses the question how quantum dynamics can be stored into quantum states and later retrieved \cite{VMC1,bisilearn,ppsr1}.
In our previous paper \cite{ppsr1} we investigated probabilistic storage and retrieval of unitary transformations, which allowed us to find covariant probabilistic universal quantum processors with exponentially smaller program register than those known before. In this paper we question how the situation changes if different prior knowledge is availale. Also we aim to present a practical description of how such probabilistic processors can be implemented in practice using elementary quantum gates.

Let us recall the formulation of the storage and retrieval task.
Consider a set of unitary channels on the $d$ dimensional Hilbert space $\hilb{H}$. Suppose one of these channels, further denoted as $\chu$,
is chosen uniformly randomly and we have access only to $N$ uses of it today. Our aim is to propose a strategy that contains channel $\chu$ $N$-times and stores it in a state of a quantum memory. This part of the task is called storage. Later, after we lost access to $\chu$, we are requested to apply $\chu$ on an unknown state $\xi$. Our goal is to choose storage and retrieval procedure in such a way that we would be able to retrieve the action of the channel $\chu$ on any state $\xi$.  The no-programming theorem implies the retrieval phase cannot be perfect universally. The approximative universality was first analyzed by
Bisio et.al. \cite{bisilearn}, where it was termed quantum learning. The perfect probabilistic version of the problem, termed \emph{probabilistic storage and retrieval of a unitary channel}(PSR) was investigated by Sedl\'ak et. al. in Ref.~\cite{ppsr1}. In this case the goal is to retrieve the quantum channel from the quantum memory only without error and with highest possible probability, which was found to be $\lambda=N/(N-1+d^2)$. The retrieval probability is required to be the same for all the unitary channels $\chu$, i.e. $\lambda=Tr(\chu(\xi))$ $\forall \chu$.

In this paper we study how this optimal success probability changes if we have some nontrivial a priori information about the unitary transformation to be stored. In particular, we will study probabilistic storage and retrieval of qubit phase gates (initiated in \cite{VMC1}), i.e. qubit unitary transformations, which in computation basis acts as
\begin{align}
\label{def:ufi}
U_\varphi=\ket{0}\bra{0}+e^{i\varphi}\ket{1}\bra{1}.
\end{align}
After finding the optimal success probability and the description of the protocol on the abstract level, we will also search for some efficient realization of the PSR protocol in terms of quantum circuit model.

The rest of the paper is organized as follows.
We use formalism of quantum combs in section \ref{sec:optim-perf-learn} for derivation of the optimal \tcb{success probability and mathematical description of the optimal protocol of $N \rightarrow 1$ PSR of qubit phase gates. In the following sections we discuss various realizations of this protocol, however, the mathematical details of Section~ \ref{sec:optim-perf-learn} are not needed for their understanding. In particular, Section} \ref{sec:1-1PSRpg} shows how a single use of a phase gate can be optimally stored and retrieved using just a single qubit storage, one CNOT gate and a single qubit measurement for retrieval.
Section \ref{sec:N-1PSRpg} gathers observations from previous two sections to describe circuit realization of the optimal $N\rightarrow 1$ protocol via an ancillary qudit and controlled shift gate followed be a measurement of the qudit.
Section \ref{sec:2-1PSRpg} specializes on $2\rightarrow1$ PSR of phase gates, i.e. a case, where the gate can be accessed twice in the storage phase. For this we minimized the CNOT gate count by hand and we present a $3$-qubit quantum circuit containing $8$ CNOTs.
Finally, in section \ref{sec:2k-1PSRpg} we show that proposal of Vidal, Masanes and Cirac \cite{VMC1} for realization of programmable phase gate in fact constitutes $(2^k-1) \rightarrow 1$ PSR of phase gates in such a way that it performs optimally and requires only $k$ CNOT gates, while having exponentially small failure probability in $k$.

\section{Optimal Probabilistic Storage and Retrieval of phase gates}
\label{sec:optim-perf-learn}

In contrast to Ref.~\cite{ppsr1} we assume here that the stored unitary channel is known to be one of the phase gates $U_\varphi$, hence, an element of $U(1)$ subgroup rather than the whole group $U(2)$ of all qubit unitary gates. In what follows we will follow conceptually the steps of \cite{ppsr1}, however, the structure of irreducible subspaces is different. Effectively, this paves the way for an increase in success probability. In this section we prove the following theorem.


\begin{theorem}
\label{thm:mainclaim}
The optimal probability of success of $N\to 1$ probabilistic storage
and retrieval of an unknown qubit phase gate $U_\varphi$ is 
$p_{\rm success}=N/(N+1)$.
\end{theorem}

\noindent
{\bf Proof}

The whole storage and retrieval protocol can be described as follows.
In the storing phase we use the $N$ copies of the unknown $U_\varphi$ to produce some state
$\ket{\psi_{\varphi}} \in \hilb{H}_M$. During the retrieving phase both the state $\ket{\psi_{\varphi}}$ and the
target state $\xi$ are sent as inputs to a retrieving quantum instrument $\mathbf{R}=\{\mathcal{R}_s, \mathcal{R}_f \}$
whose output in the case of successful retrieving ($\mathcal{R}_s$) should be exactly $\mathcal{U_\varphi}(\xi)=U_\varphi(\xi)U_\varphi^\dagger$.
  Any possible way in which storing and retrieving can be done (parallel or sequential application of $U_\varphi$, or any other intermediate approach) is mathematically  described by
inserting the $N$ uses of the unitary channel $\mathcal{U_\varphi}$ into $N$ open slots of a generalized quantum instrument (see supplementary material of \cite{ppsr1} for a short review, or \cite{comblong,architecture,supermaps})  $\mathbf{L}=\{\mathcal{L}_s, \mathcal{L}_f \}$:

\begin{align}
  \begin{aligned}
\mathbf{L}
\qquad \qquad \qquad
\qquad \;\:\, \\[-4pt]
\overbrace{\qquad \qquad \qquad \qquad \qquad \qquad \qquad \qquad
\quad \;\;\;}\\
\Qcircuit @C=0.4em @R=1em
 {
\multiprepareC{1}{\,}&
\ustick{\scriptstyle{1}} \qw &
\gate{U_\varphi}&
\ustick{\scriptstyle{2}}\qw&
\multigate{1}{\,\,\,}&
\ustick{\scriptstyle{3} }\qw&
\gate{U_\varphi}&
\ustick{\scriptstyle{4} }\qw&
\pureghost{\dots}&
&
\ustick{\scriptstyle{2N}}\qw &
\multigate{1}{\,\,\,}&
&
&
&
\ustick{\scriptstyle{\!\!\!2N+1}}&
\multigate{1}{\,\,\,}&
\ustick{\scriptstyle{\;\;\; 2N+2}}\qw
\\
\pureghost{\,}&
\qw&
\qw&
\qw&
\ghost{\,\,\,}&
\qw&
\qw&
\qw&
\cdots&
&
\qw &
\ghost{\,\,\,}&
\qw&
\ustick{\;\scriptstyle{M}}\qw&
\qw&
\qw&
\ghost{\,\,\,}&
}\\[-8pt]
\underbrace{\qquad \qquad \qquad \qquad \qquad \qquad \qquad \;}
\quad \underbrace{\qquad \;}
\\[-4pt]
\mathcal{S}
\qquad \qquad \qquad \qquad \;\:\: \mathbf{R} \; \; \,
\label{eq:learnnetwork}
  \end{aligned}
\end{align}
where
$\mathcal{S}$
denotes the (deterministic) storing network and
$\mathbf{R}$ the retrieving quantum instrument.
 The output system of the storing network
corresponds to the Hilbert space
$\hilb{H}_M$ which carries the state
$\ket{\psi_{\varphi}}$.
In the case of successful retrieving (i.e. observing outcome $s$ corresponding to both $\mathcal{R}_s$ and $\mathcal{L}_s$) the resulting quantum operation from $\mathcal{L}(\hilb{H}_{2N+1})$ to
 $\mathcal{L}(\hilb{H}_{2N+2})$ is required to be proportional to
 the channel $\mathcal{U}_\varphi$ i.e.
 \begin{align}
 \label{eg:perfectcond1}
   L_s *
\left(
\bigotimes_{i=1}^{N}
\KetBra{U_\varphi}{U_\varphi}_{2i-1,2i}
 \right) = \lambda
\KetBra{U_\varphi}{U_\varphi}_{2N+1,2N+2},
 \end{align}
where we used the link product formalism
and the Choi operator $L_s = S*R_s$ describes the successful operation of the storing and retrieving quantum network.
Thus, in this case we know with certainty
that final output of the network is the desired state
$U_\varphi \xi U_\varphi^\dag \in \mathcal{L}(\hilb{H}_{2N+2})$.
By expressing the link product in the above Eq. (\ref{eg:perfectcond1}) explicitly
the requirement of perfect probabilistic storing and retrieving can be stated as
\begin{align}
  \label{eq:perfectlearcond}
\Bra{U_\varphi^*}^{\otimes N}
L_s
\Ket{U_\varphi^*}^{\otimes N}
=
\lambda \KetBra{U_\varphi}{U_\varphi} \quad\quad \forall \varphi \in [0,2\pi].
\end{align}
We stress that the probability of success, i.e. the value of $\lambda$ is required to be the same for all $\varphi \in [0,2\pi]$.
The aim of our analysis is to derive the optimal probabilistic quantum network $L_s$, which obeys the constraint of Eq. \eqref{eq:perfectlearcond}
and maximizes the value of $\lambda$.

Our first observation is that the operator
$L_s$ could be chosen to satisfy the commutation relation
\begin{align}
  \label{eq:comrel}
  [ L_s,  U_\varphi^{\otimes N} \otimes U_\vartheta^{\otimes N} \otimes  {(U_\varphi^*)_{2N+1}} \otimes
  (U_\vartheta^*)_{2N+2}] = 0
\end{align}
for all $\varphi,\vartheta \in [0,2\pi]$.
This can be proven by showing that any optimal strategy can be made covariant, while keeping the same success probability.
As a consequence of Eq. \eqref{eq:comrel}, it was proved in \cite{bisilearn} that the optimal storing phase
is \emph{parallel}, i.e. the $N$ uses of the unknown
unitary are applied in parallel on a quantum state $\ket{\psi}$ as shown in
the following diagram:

  \begin{align}
    \begin{aligned}
      \Qcircuit @C=0.4em @R=0.5em
 {
\prepareC{\psi_\varphi}&\qw
}
    \end{aligned}
=
  \begin{aligned}
\Qcircuit @C=0.4em @R=0.3em
 {
\multiprepareC{3}{\psi}&
\ustick{\scriptstyle{1}} \qw &
\gate{U_\varphi}&
\ustick{\scriptstyle{2}}\qw&
\multigate{3}{\,\,\,}&
&
\\
\pureghost{\psi}&
\ustick{\scriptstyle{3}} \qw &
\gate{U_\varphi}&
\ustick{\scriptstyle{4}} \qw &
\ghost{1}{\,\,\,}&
&
\\
\pureghost{\psi}&
&
\vdots&
&
\pureghost{1}{\,\,\,}&
&
\\
\pureghost{\psi}&
\qw&
\qw&
\qw&
\ghost{\,\,\,}&
\ustick{\;\scriptstyle{M}}\qw&
}
  \end{aligned}
=
  \begin{aligned}
\Qcircuit @C=0.4em @R=1em
 {
\multiprepareC{1}{\psi}&
\ustick{\scriptstyle{A}} \qw &
\gate{U_\varphi^{\otimes N}}&
\ustick{\scriptstyle{B}}\qw&
\multigate{1}{\,\,\,}&
&
\\
\pureghost{\psi}&
\qw&
\ustick{\scriptstyle{A'}}\qw&
\qw&
\ghost{\,\,\,}&
\ustick{\scriptstyle{\;M}}\qw&
}
  \end{aligned}\;.
\label{eq:parallelstor}
\end{align}
In this diagram we use labels $A, B$ to denote all input, output Hilbert spaces of $N$ uses of the phase gate, respectively.

Let us now consider the decomposition of
$U_\varphi^{\otimes N} \in \mathcal{L}(\hilb{H}_A)$ into irreducible representations (irreps) of $U(1)$
\begin{align}
  \label{eq:irrepsUN}
  U_\varphi^{\otimes N} = \bigoplus_{j=0}^{N} e^{ij\varphi} \otimes I_{m_j},
\end{align}
where $I_{m_j}$ denotes the identity operator on the multiplicity space. Let us remind that all irreps of $U(1)$ are one dimensional ($\dim(\hilb{H}_j) = 1$) and $e^{ij\varphi}$ represents the element $e^{i\varphi} \in U(1)$.
Eq. \eqref{eq:irrepsUN} induces the following decomposition of the Hilbert
space $\hilb{H}_A$
\begin{align}
  \label{eq:decompohilb}
   \hilb{H}_A := \bigoplus_j \hilb{H}_j \otimes \hilb{H}_{m_j}  \quad\quad
 \quad  \dim(\hilb{H}_{m_j}) = m_j .
\end{align}
It was shown in \cite{bisilearn}
that the optimal state $\ket{\psi}$ for the storage
can be taken of the following form
\begin{align}
  \label{eq:optstate}
  \begin{aligned}
  \ket{\psi} :=  \bigoplus_j \sqrt{p_j} \Ket{I_j} \in
  \tilde{\mathcal{H}}\quad\quad
p_j \geq 0, \; \sum_j p_j =1
  \end{aligned}
\end{align}
where $\hilb{H}_A\otimes \hilb{H}_{A'} \supseteq \tilde{\hilb{H}} :=
\bigoplus_j \hilb{H}_j\otimes \hilb{H}_j$ and
$I_j$ denotes the identity operator on $\hilb{H}_j$.
The  optimal state $\ket{\psi} $ undergoes the action of the unitary
channels and becomes
$\ket{\psi_\varphi} :=\bigoplus_j \sqrt{p_j} e^{ij\varphi}\Ket{I_j} $.
Clearly, $\ket{\psi_\varphi}$ belongs to $\hilb{H}_M$ which is a subspace of
$\hilb{H}_B\otimes \hilb{H}_{A'}$  isomorphic
to $\tilde{\hilb{H}} $.

We can focus our attention on the retrieving quantum instrument
$\{ \mathcal{R}_s, \mathcal{R}_f \}$ from $\mathcal{L}(\hilb{H}_C\otimes
\hilb{H}_M)$
to
$\mathcal{L}(\hilb{H}_D)$
  \begin{align}
  \begin{aligned}
\Qcircuit @C=1em @R=1em
 {
 \ustick{\scriptstyle{C}}
&
\multigate{1}{\mathcal{R}_{i=r,s}}&
\ustick{\scriptstyle{D}} \qw &
\\
\ustick{\scriptstyle{M}}
&
\ghost{\mathcal{R}_{i=r,s}}&
&
}
  \end{aligned}\;.
\label{eq:retrievinstr}
\end{align}
The condition that the outcome $s$ corresponds to the
perfect learning becomes:
  \begin{align}
R_s * \ketbra{\psi_{\varphi}}{\psi_{\varphi}} &= \Tr_M[R_s
    ((\ketbra{\psi_\varphi}{\psi_\varphi})^T\otimes I_{C,D}) ] \nonumber \\
    &=\bra{\psi^*_\varphi} R_s \ket{\psi^*_\varphi} 
    = \lambda \KetBra{U_\varphi}{U_\varphi}\quad \forall \varphi\in[0,2\pi]
\end{align}

 \begin{align}
&\begin{aligned}
\Qcircuit @C=0.7em @R=1em
 {
&
 \ustick{\scriptstyle{A}}
&
\multigate{1}{\mathcal{R}_{s}}&
\ustick{\scriptstyle{D}} \qw &
\\
 \prepareC{\psi_\varphi}&
\ustick{\scriptstyle{M}} \qw
&
\ghost{\mathcal{R}_{s}}&
&
}
  \end{aligned}
\;\; = \;\;\lambda \; \;
 \Qcircuit @C=0.6em @R=0.6em
  {
 &\gate{U_\varphi}& \qw
 } \quad,
\label{eq:perfectretriev}
\end{align}
where $\ket{\psi^*_\varphi} = \bigoplus_j \sqrt{p_j}
e^{-ij\varphi}\Ket{I_j}$.
The optimal $R_s$ can be chosen to satisfy the following commutation relation:
\begin{align}
  \label{eq:commretriev}
&    \left [ R_s,U_\varphi' U_\vartheta' \otimes (U_\varphi^*)_C\otimes (U_\vartheta)^*_D   \right ]=0, \\
& \qquad \quad U_{\varphi}' := \bigoplus_j e^{ij\varphi}\; I_{j} \otimes I_{j}. \nonumber
\end{align}
which is clearly the analog of Eq.~\eqref{eq:comrel} where
$U_\varphi^{\otimes N} \otimes U_\vartheta^{\otimes N}$ has been replaced by $U_\varphi' U_\vartheta'$.
Then, reminding that $U_\varphi'\ket{\psi}= \ket{\psi_\varphi}$ and
$\ket{\psi^*}=\ket{\psi}$, from Eq.~\eqref{eq:commretriev}
we have
\begin{align}
  \label{eq:simplifiedlambda}
  \begin{aligned}
   & \bra{\psi^*_\varphi} R_s \ket{\psi^*_\varphi} = \lambda \KetBra{U_\varphi}{U_\varphi} \quad
\forall \varphi \iff
    \bra{\psi} R_s \ket{\psi} = \lambda \KetBra{I}{I} \; .
  \end{aligned}
\end{align}

Let us now summarize what we discussed so far by giving a formal
statement of the optimization problem for
probabilistic storage and retrieval of a qubit phase gate:
\begin{align}
  \label{eq:optimization}
  \begin{aligned}
    & \underset{\ket{\psi},R_s}{\mbox{\rm maximize}}
& & \lambda =
\frac{1}{4}\Bra{I}\bra{\psi} R_s \ket{\psi} \Ket{I}
 \\
& \mbox{\rm subject to}
&& \bra{\psi}R_s\ket{\psi} = \lambda \KetBra{I}{I}\\
&& & \ket{\psi} \mbox{ as in Eq. \eqref{eq:optstate} }\\
&&& R_s   \mbox{ obeys Eq. \eqref{eq:commretriev} }\\
&&& \Tr_D[R_s] \leq I \; .
  \end{aligned}
\end{align}

Consider now the decomposition
\begin{align}
  \label{eq:decompopartial}
  \begin{aligned}
    e^{ij\varphi}I_j \otimes U^*_\varphi = \bigoplus_{J\in \mathsf{J}_j} e^{iJ\varphi}I_J \otimes I_{m^{(j)}_{J}} \\
    \hilb{H}_j \otimes \hilb{H} = \bigoplus_{J\in \mathsf{J}_j}
    \hilb{H}_J \otimes \hilb{H}_{m^{(j)}_{J}},
\end{aligned}
\end{align}
where the index $j$ labels the irreducible representations
in the decomposition of $U_\varphi^{\otimes N}$ and we denote with
$\mathsf{J}_j$ the set of values of $J$ such that $e^{iJ\varphi}$ 
is in the decomposition of $e^{ij\varphi}I_j \otimes U^*_\varphi$.
It is important to notice that the multiplicity spaces
$\hilb{H}_{m^{(j)}_{J}} $  are one dimensional and therefore
$ I_{m^{(j)}_{J}}$ are rank one.
Then we have
\begin{align}
  \label{eq:decompototal}
  \begin{aligned}
    U'_\varphi U'_\vartheta \otimes U_\varphi^*\otimes U_\vartheta^* =
    \bigoplus_{J,K=-1}^N e^{iJ\varphi} I_J\otimes e^{iK\vartheta}I_K \otimes I_{m_{JK}}  \\
    \hilb{H}_{m_{JK}} = \bigoplus_{j\in \mathsf{j}_{JK}}
    \hilb{H}_{m_J^{(j)}}\otimes\hilb{H}_{m_K^{(j)}}
  \end{aligned}
\end{align}
where $\mathsf{j}_{JK}$ denotes the set of values of
$j$ such that $ e^{iJ\varphi}e^{iK\vartheta}$ is in the decomposition of
$e^{ij\varphi}I_j \otimes e^{ij\vartheta}I_j \otimes U^*_\varphi \otimes U^*_\vartheta $.
For example, for $J=K=0,\ldots,N-1$ $\mathsf{j}_{JJ}=\{J,J+1\}$, $\mathsf{j}_{-1\;-1}=\{0\}$, $\mathsf{j}_{NN}=\{N\}$. Since $\dim(\hilb{H}_{m_J^{(j)}})=1$
we stress that $\BraKet{I_{m_J^{(j)}}}{I_{m_J^{(j')}}}=\delta_{j,j'}$, $\ket{\chi} \in
\hilb{H}_{m_J^{(j)}}\otimes\hilb{H}_{m_J^{(j)}}$ $\Leftrightarrow $ $\ket{\chi} \propto \Ket{I_{m_J^{(j)}}} $
and $ \hilb{H}_{m_{JJ}} = \spn(\{\Ket{I_{m_J^{(j)}}}\}, j\in \mathsf{j}_{JJ})$.

From Eq.~\eqref{eq:decompototal} the commutation relation of
Eq.~\eqref{eq:commretriev}
becomes
\begin{align}
  \label{eq:comretriev2}
  \begin{aligned}
    \left[
R_s ,  \bigoplus_{J,K=-1}^N \; e^{iJ\varphi} I_J\otimes e^{iK\vartheta}I_K \otimes I_{m_{JK}}
\right]=0
  \end{aligned}
\end{align}
which, thanks to the Schur's lemma, gives
\begin{align}
  \label{eq:decompor}
  \begin{aligned}
    R_s = \bigoplus_{J,K=-1}^N I_J \otimes I_K \otimes s^{(JK)} \\
s^{(JK)} \in \mathcal{L}(\hilb{H}_{m_{JK}})\;\;, s^{(JK)}\geq0
  \end{aligned}
\end{align}
From Eq. \eqref{eq:decompor} we have that the quantum operation
$R_s$ is the sum of the positive operators
$ I_J \otimes I_K \otimes s^{(JK)}$.
Therefore we have that
\begin{align}
\label{eq:perfectlearcondJJ}
\bra{\psi}  R_s \ket{\psi} &= \lambda \KetBra{I}{I} \iff    \\
&\bra{\psi}   I_J \otimes I_K \otimes s^{(JK)} \ket{\psi} =
\lambda_{JK} \KetBra{I}{I} \quad \forall J,K
\nonumber
\end{align}
since $ \KetBra{I}{I}$ is a rank one operator.

From the identity 
$I_j \otimes I = \bigoplus_{J=j-1}^j I_J \otimes I_{m^{(j)}_{J}}$ (we remind that $ I_{m^{(j)}_{J}}$
has rank one), we obtain
\begin{align}
  \label{eq:relevantstate}
      \ket{\psi}\Ket{I} &= \bigoplus_{j=0}^N \bigoplus_{J=j-1}^j
\sqrt{p_j} \Ket{I_J}\Ket{I_{m_J^{(j)}}} \\
&= \bigoplus_{J=-1}^N \bigoplus_{j\in \mathsf{j}_{JJ}} 
\sqrt{p_j} \Ket{I_J}\Ket{I_{m_J^{(j)}}} 
= \bigoplus_{J=-1}^N \Ket{I_J} \ket{\phi_J} \nonumber\\
\ket{\phi_J}&:= \bigoplus_{j\in \mathsf{j}_{JJ}}
\sqrt{p_j} \Ket{I_{m_J^{(j)}}}.
 \end{align}

Using Eq.~\eqref{eq:decompor} into Eq.~\eqref{eq:optimization}
we obtain
\begin{align}
  \lambda_{JK}&= \delta_{JK}\lambda_{J},\qquad  \lambda = \sum_{J=-1}^N \lambda_{J} \\
\lambda_{J} &= \frac{1}{4} \bra{\phi_J}s^{(JJ)}\ket{\phi_J}
\label{eq:lambdaJ}
 \end{align}
where the
$\lambda_{JK}$'s were defined in Eq. (\ref{eq:perfectlearcondJJ}). 
It is now easy to show that we can assume
\begin{align}
  \label{eq:diagonalR}
    R_s &= \bigoplus_{J} I_J \otimes I_J \otimes s^{(J)} \qquad \\
s^{(J)} &:= \sum_{j,j' \in \mathsf{j}_{JJ}} s^{(J)}_{jj'} \KetBra{I_{m_J^{(j)}}}{I_{m_J^{(j')}}}
\end{align}
Indeed, let $R_s'=\bigoplus_{JK} I_J \otimes I_K \otimes s'^{(JK)}$ be
the optimal retrieving quantum operation and let us define linear operators
$R_s=\bigoplus_{J} I_J \otimes I_J \otimes s^{(J)}$
where $s^{(J)} =  s'^{(JJ)}$ and
 $R''_s=\bigoplus_{J\neq K} I_J \otimes I_K \otimes s^{(JK)}$.
Since both $R_s$ and $R_s''$ are positive and
$R_s+R_s'' = R_s'$, we have that normalization condition $\Tr_D[R_s']\leq I$ implies
 $\Tr_D[R_s]\leq I$, $\Tr_D[R_s'']\leq I$ i.e. $R_s$ and $R_s''$ are quantum operations.
Finally, from Eq.~\eqref{eq:lambdaJ} we have that
 $\bra{\psi}R_s\ket{\psi} = \bra{\psi}R'_s\ket{\psi}$, thus proving
 that also $\{R_s,\ket{\psi} \}$ is an optimal solution of the
 optimization problem \eqref{eq:optimization}.

If $R_s$ is of the form of
  Eq.~\eqref{eq:diagonalR} we can express the constraint 
of Eq. (\ref{eq:simplifiedlambda}) in terms of the operators
$s^{(J)}$ as follows:
\begin{align}
s^{(J)}_{j,j'}=\frac{\mu_J}{\sqrt{p_j p_{j'}}} \quad J=0,\ldots,N-1, \quad s^{(-1)}=s^{(N)}=0,
\label{eq:reformperfectlearn}
\end{align}
where $\mu_J$ is some number, which must be non-negative due to positive-semidefinitness of $R_S$.
The proof of Eq.~\eqref{eq:reformperfectlearn} is given in appendix
\ref{sec:proof-oflemma1}.

Fulfillment of Eq. (\ref{eq:reformperfectlearn}) guarantees the perfect storing and retrieving of phase gates and we can rewrite the probability of success as
\begin{align}
  \label{eq:19}
  \lambda &
  =\sum_{J=0}^{N-1} \frac{1}{4}\!\!\! \sum_{j,j' \in \mathsf{j}_{JJ} }\sqrt{p_j} \frac{\mu_{J}}{\sqrt{p_j p_{j'}}} \sqrt{p_{j'}}
  =   \sum_{J=0}^{N-1} \mu_{J},
\end{align}
where we used Eqs. (\ref{eq:lambdaJ}), (\ref{eq:reformperfectlearn}).

Let us now consider the trace non-increasing constraint for the retrieving
quantum operation, which for the Choi operator reads
$\Tr_D[R_s]\leq I$. Since $R_s$ satisfies Eq.~\eqref{eq:commretriev},
we have that $ \left[  \Tr_D[R_s],U_\varphi' U_\vartheta' \otimes (U_\varphi^*)_C\right]=0$ implies
\begin{align}
  \label{eq:partialtrace}
\Tr_D[R_s] = \bigoplus_{J=-1}^N\bigoplus_{j\in \mathsf{j}_{JJ}} I_J \otimes
I_j \, s^{(J)}_{jj}
\end{align}
From Eq.~\eqref{eq:partialtrace} we have
\begin{align}
  \label{eq:subnormalization}
  \begin{aligned}
    \Tr_D[R_s]\leq I \Leftrightarrow
s^{(J)}_{jj} \leq 1 \; \; J=-1,\dots,N, \; \forall j\in  \mathsf{j}_{JJ}.
  \end{aligned}
\end{align}
Thanks to Eq. (\ref{eq:reformperfectlearn}) the above can be expressed as inequalities between $\mu_J$ and $p_j$ as
\begin{align}
  \label{eq:20}
  \mu_J \leq p_j \quad \forall j \in \mathsf{j}_{JJ}  \;\;\;J=-1,\ldots,N
\end{align}

Collecting Eqs.(\ref{eq:19}),(\ref{eq:20}) and (\ref{eq:optstate}) the optimization of perfect probabilistic storing and retrieving can be reduced to
\begin{align}
  \label{eq:optimizationpjmj}
    & \underset{\mu_J, p_j}{\mbox{\rm maximize}}
& & \lambda =\sum_{J=0}^{N-1} \mu_{J},
\\
& \mbox{\rm subject to}
&&0 \leq \mu_J  \leq p_j\quad \forall j \in \mathsf{j}_{JJ}  \;\;\;J=0,\ldots,N-1  \nonumber \\
&&& p_j \geq 0 \quad \sum_j p_j=1, \nonumber
\end{align}

Let us write inequalities that are given by Eq. (\ref{eq:20}). For any $J=0,\ldots,N$ we have
\begin{align}
  \label{eq:ineq1}
  \mu_{J}\leq p_J \\
  \label{eq:ineq2}
  \mu_{J-1}\; \leq p_{J}
\end{align}
We define nonnegative coefficient $f_J\in [0,1]$ for $J=0, \ldots ,N$ via the formula $f_J=(N-J)/N$.
We can multiply Eq. (\ref{eq:ineq1}) by $f_J$, Eq. (\ref{eq:ineq2}) by $1-f_J$ and  sum them up for all $J$. We obtain
\begin{align}
\label{eq:firstub0}
\sum_{J=0}^{N} f_J\; \mu_{J}\;+ (1-f_{J}) \;\; \mu_{J-1}\;\leq \sum_{J=0}^{N} p_J=1
\end{align}
The above inequality can be rewritten as $\sum_{J=0}^{N-1} \frac{N+1}{N} \mu_{J}  \leq 1,$
which proves that $\lambda\leq N/(N+1)$.

Let us mention that the coefficient $f_J$ was intentionally chosen so that the coefficient multiplying $\mu_J$ is constant and we get an upper bound on $\lambda$ in Eq. (\ref{eq:optimizationpjmj}).

Finally, we finish the proof of Theorem 1 by showing that the obtained upper bound can be saturated. One can simply choose
%
\begin{align}
p_j&=\frac{1}{N+1} \;\;\;j=0,\ldots,N  \nonumber\\
\mu_J&=\frac{1}{N+1} \;\;\; J=0,\ldots,N-1
\end{align}
and check that conditions in Eq. (\ref{eq:optimizationpjmj}) are satisfied and $\lambda=N/(N+1)$. Knowledge of $\mu_J$ and $p_j$ allows us to completely specify the state $\ket{\psi}$ and the retrieving operation $\mathcal{R}_s$ sufficient for building the complete storing and retrieving strategy.
\qed

\begin{figure}
  \begin{center}
    \includegraphics[width=8cm]{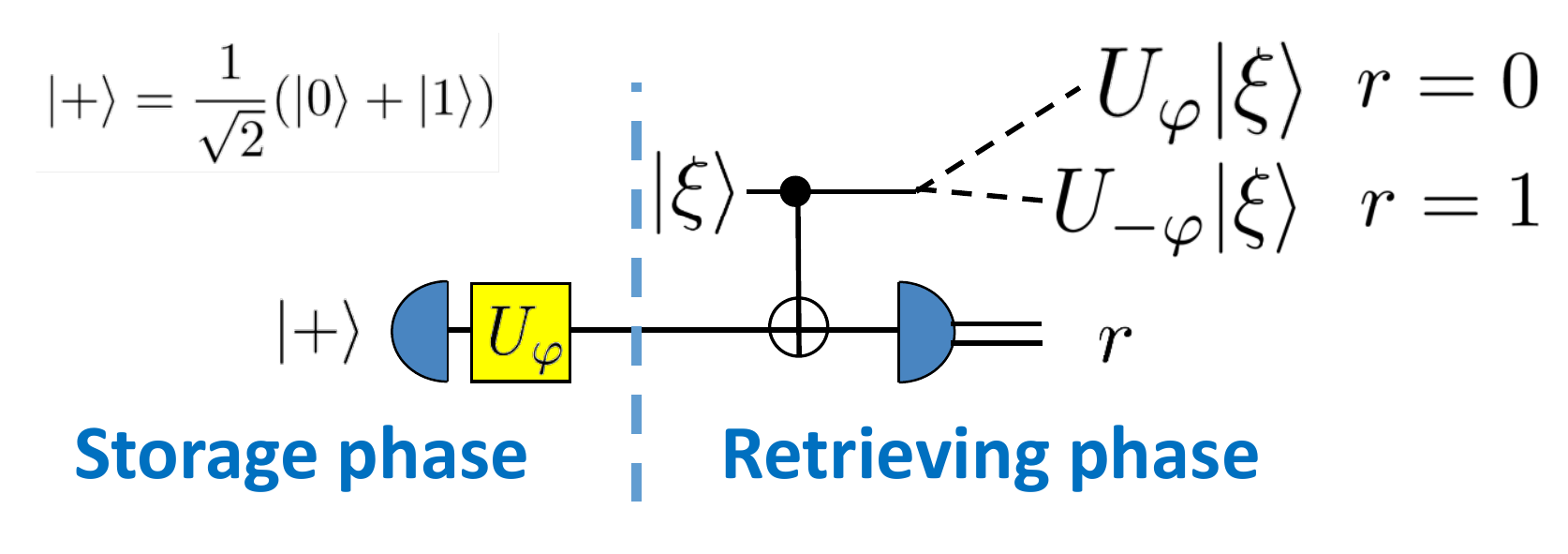}
\caption{Optimal $1\rightarrow 1$ PSR of phase gates.}
\label{fig:1na1}
  \end{center}
\end{figure}

\section{$1\rightarrow 1$ Optimal circuit realization}
\label{sec:1-1PSRpg}
  It follows from Theorem~\ref{thm:mainclaim} that in case of one to one
storing and retrieval the success probability equals $p_{\rm success}=1/2$.
Such success probability can be easily obtained (see Fig. (\ref{fig:1na1})) if we apply the gate on a state $\ket{+}=(\ket{0}+\ket{1})/\sqrt{2}$ and feed the resulting state $\ket{\psi_\varphi}=(\ket{0}+e^{i \varphi}\ket{1})/\sqrt{2}$ as a program state into CNOT gate.
Unitary operator $C_{\oplus}$ representing the CNOT gate acts on two qubits as $\ket{j}\otimes\ket{k}\mapsto\ket{j}\otimes\ket{j\oplus k}$. We usually say that the first 
qubit is the control qubit and the second one is called the target qubit.

Properties of the CNOT gate $C_\oplus$ as a probabilistic programmable processor were studied by Vidal, Masanes and Cirac \cite{VMC1}, who exploited this device for probabilistic  programmable implementation of phase gates. In particular, assume the control qubit is initialized in state $\ket{\xi}=\alpha \ket{0} + \beta \ket{1}$ and the target qubit stores the action of $U_\varphi$, i.e. it is in the state $\ket{\psi_\varphi}$. Then
\begin{align}
\label{eq:CNOT11}
C_{\oplus} (\ket{\xi}\otimes \ket{\psi_\varphi} ) &\mapsto
\frac{1}{\sqrt{2}}(\alpha\ket{00}+ \alpha e^{i \varphi} \ket{01} + \beta \ket{11} + \beta e^{i \varphi} \ket{10}  \nonumber \\
&=U_{\varphi}\ket{\xi}\otimes  \frac{1}{\sqrt{2}}\ket{0}+U_{-\varphi}\ket{\xi}\otimes \frac{1}{\sqrt{2}}\ket{1}\,.
\end{align}
We see that measurement of the second (target) qubit in the computational basis will yield both outcomes with the same probability 1/2. Outcome 0 indicates the implementation of $U_\varphi$ was successful, because the first qubit is collapsed into the desired state $U_{\varphi}\ket{\xi}$. Otherwise, the resulting state is rotated in the opposite direction $U_{-\varphi}\ket{\xi}$.

\section{$N\rightarrow 1$ Realization via ancilary qudit}
\label{sec:N-1PSRpg}
The aim of this section is to generalize the construction presented in the previous section for general $N$. Thanks to $U(1)$ irreps being one-dimensional it is not necessary to use ancillary system in parallel with the unknown gate $U_\varphi$ for the input state $\ket{\psi}$.
Thus, instead of taking $\ket{\psi}$ as in Eq. (\ref{eq:optstate}) we can take $\ket{\psi} =  \sum_{j=0}^{N} \frac{1}{\sqrt{N+1}} \ket{v_j} \in \hilb{H}_A$, where $\ket{v_j}$ is any normalized vector defining an irrep $e^{ij\varphi}$ of $U(1)$ in $\hilb{H}_A$, i.e. any element of the computational basis with $j$ ones.  The reason is that a fixed unitary can interlink those two $(N+1)-$dimensional subspaces and the value of the amplitudes $\sqrt{p_j}=1/ \sqrt{N+1}$ follows from proof of Theorem \ref{thm:mainclaim}.

Let us define a virtual qudit ($D$ dimensional Hilbert space)
identified with the subspace $V_D={span}\{\ket{v_j}\}_{j=0}^N$
of dimension $D=N+1$ and denote its basis states as
$\{\ket{t}\equiv \ket{v_t}\}_{t=0}^{N}$.
We denote $P_D=\sum_{t=0}^{N} \ket{t}\bra{t}$ the projector onto $V_D$, and by $P^\perp_D=I-P_D$ the projector onto its orthocomplement.
During the storage
$\ket{\psi}\in V_D\subset\hilb{H}_A$ evolves into $\ket{\psi_\varphi} = \frac{1}{\sqrt{N+1}} \sum_{j=0}^{N} e^{ij\varphi} \ket{v_j} \in V_D$.
We can now define a channel $\mathcal{E}$, which maps states from $\hilb{H}_A$ to states on $V_D$ and on subspace $V_D$ acts as identity. This is achieved by
\begin{align}
\label{eq:channelE}
\mathcal{E}(\rho)=P_D\;\rho\;P_D + Tr(\rho P^\perp_D) \ket{t_0}\bra{t_0},
\end{align}
where $\rho\in \mathcal{L}(\hilb{H}_A)$ and $\ket{t_0}$ is some state in $V_D$.
In particular, $ Tr(\ket{\psi_\varphi} \bra{\psi_\varphi}  P^\perp_D)=0$,
thus $\mathcal{E}(\ket{\psi_\varphi} \bra{\psi_\varphi})=\ket{\psi_\varphi} \bra{\psi_\varphi}$, since $\ket{\psi_\varphi}\in V_D$.

Next, we define \emph{a control shift-down gate $C_{\ominus}$} as a bipartite gate with control qubit and a target qudit via the formula
\begin{align}
\label{def:cshiftdown}
C_{\ominus}\ket{c}\otimes \ket{t} \mapsto \ket{c}\otimes \ket{t\ominus c}.
\end{align}
Suppose that the state on which the retrieved gate should act is again a pure state $\ket{\xi}=\alpha \ket{0} + \beta \ket{1}$.
The control shift-down gate $C_{\ominus}$ whose control qubit is in a state $\ket{\xi}$ and target qudit is in the state $\ket{\psi_\varphi}$ acts as
\begin{align}
\label{eq:Cshiftdown1}
C_{\ominus}& (\ket{\xi}\otimes \ket{\psi_\varphi} ) \mapsto \nonumber \\
&\alpha\ket{0}\otimes\frac{1}{\sqrt{N+1}} \sum_{t=0}^{N} e^{it\varphi} \ket{t}  \nonumber \\
&+\beta\ket{1}\otimes\frac{1}{\sqrt{N+1}} \left(\sum_{t=1}^{N} e^{it\varphi} \ket{t-1} + \ket{N}\right)  \nonumber \\
&=U_{\varphi}\ket{\xi}\otimes  \frac{1}{\sqrt{N+1}} \sum_{t=0}^{N-1}  e^{it\varphi} \ket{t} \nonumber\\
&+U_{-N\varphi}\ket{\xi}\otimes \frac{e^{iN\varphi}}{\sqrt{N+1}} \ket{N}
\end{align}

\begin{figure}
  \begin{center}
    \includegraphics[width=8.5cm]{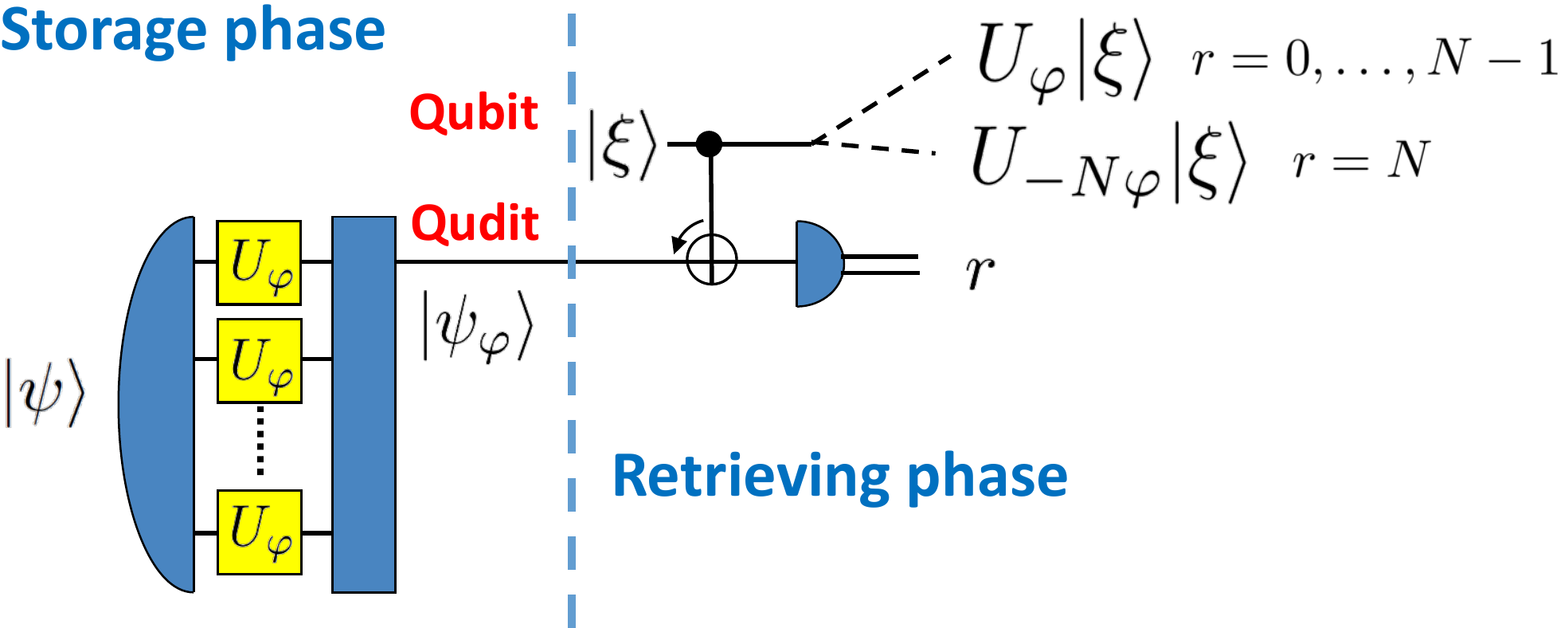}
\caption{Optimal $N\rightarrow 1$ PSR of qubit phase gates using single qudit and one control shift down gate (see text for details).}
\label{fig:quditNna1}
  \end{center}
\end{figure}

Last step of the implementation (see Fig. (\ref{fig:quditNna1}) for illustration) is a measurement of the qudit in its computational basis $\{\ket{t}\}_{t=0}^{N}$. The probability of observing outcome $t$ is $1/(N+1)$ and it is calculated as $Tr\left( C_{\ominus} (\ket{\xi}\bra{\xi}\otimes \mathcal{E}(\ket{\psi_\varphi}\bra{\psi_\varphi})) C_{\ominus}^\dagger \;\; I\otimes \ket{t}\bra{t} \right)$.
The post-measurement state of a qubit is in case of outcome $t=0,\ldots,N-1$ the same and reads
\begin{align}
\label{eq:postSuccess}
U_{\varphi}\ket{\xi}& \bra{\xi} U_{\varphi}^\dagger = \nonumber \\
  &= Tr_{2}\left( C_{\ominus} (\ket{\xi}\bra{\xi}\otimes \mathcal{E}(\ket{\psi_\varphi}\bra{\psi_\varphi})) C_{\ominus}^\dagger \;\; I\otimes \ket{t}\bra{t} \right),
\end{align}
while for outcome $t=N$ the qubit collapses into a state $U_{-N\varphi}\ket{\xi}\bra{\xi} U_{-N\varphi}^\dagger$.
At this point it is easy to see that the presented implementation of the storage and retrieval of the phase gate would work also for mixed input states $\xi$ due to linearity of quantum mechanics.
We conclude that the presented realization succeeds with optimal probability $N/(N+1)$, since this is the total probability of obtaining result other than $t=N$.

\section{$2\rightarrow1$ Circuit realization}
\label{sec:2-1PSRpg}

\begin{figure}
  \begin{center}
    \includegraphics[width=7cm]{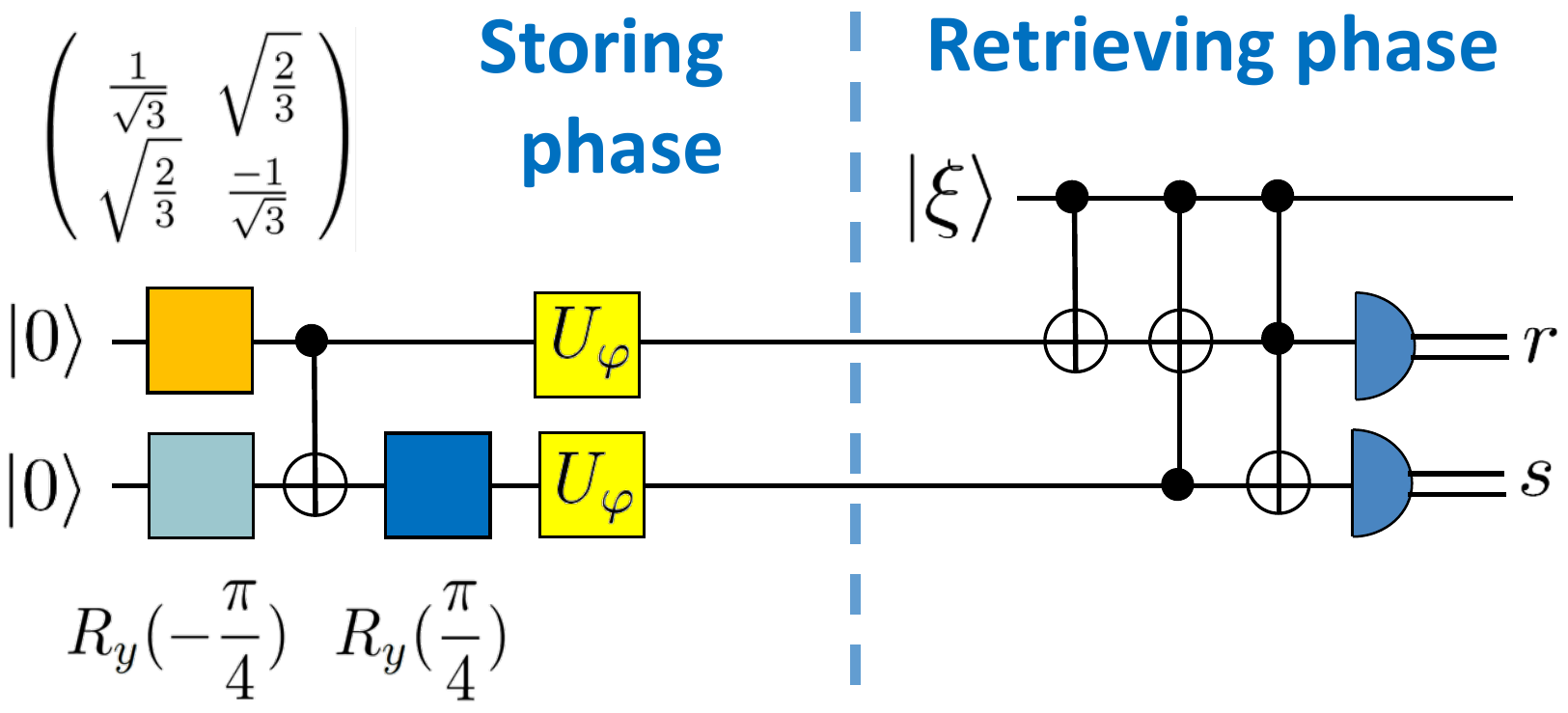}
\caption{Optimal $2\rightarrow 1$ PSR of phase gates using $2$ CNOT gates, $2$ Toffoli gates and $3$ fixed one qubit gates.}
\label{fig:2na1toffoli}
  \end{center}
\end{figure}

This section specializes on the case, where the gate $U_\varphi$ can be
accessed twice in the storage phase.
We present a design of the optimally performing circuit, which follows the ideas from the previous section, but our aim here is to decompose all the operations into elementary quantum gates \cite{barenco}. The first part of the circuit (see Fig. \ref{fig:2na1toffoli}) formed by a CNOT gate, $R_y(\pi/4)=\exp[i\frac{\pi}{8}\sigma_y]$, $R_y(-\pi/4)$ and one qubit gate
\begin{align}
M=\left(
\begin{array}{cc}
\frac{1}{\sqrt{3}}&\sqrt{\frac{2}{3}}\\
\sqrt{\frac{2}{3}}&\frac{-1}{\sqrt{3}}\\
\end{array}
\right)
\end{align}
transforms the second and third qubit from state $\ket{00}$ into state
\begin{align}
\ket{\psi} = \frac{1}{\sqrt{3}} (\ket{00}+\ket{10}+\ket{11}). \\
\end{align}
The action of the phase gate $U_\varphi$ on the second and third qubit leads to a state $\ket{\psi_\varphi} = \frac{1}{\sqrt{3}} (\ket{00}+e^{i\varphi}\ket{10}+e^{i 2 \varphi}\ket{11})$. We chose subspace $V_3=span\{\ket{00},\ket{10},\ket{11})\}$ as our virtual qutrit.
\begin{figure}
  \begin{center}
    \includegraphics[width=5cm]{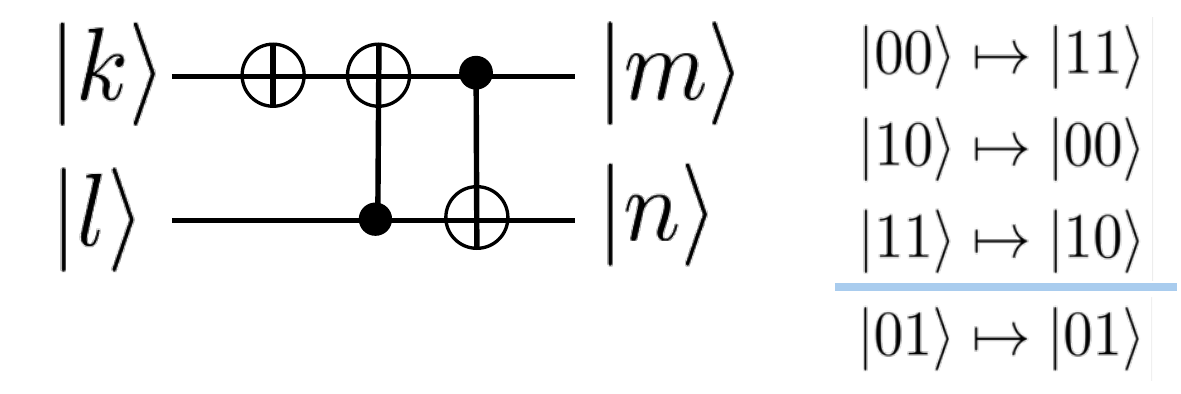}
\caption{Small quantum circuit performing shift down operation in the subspace $V_3=span\{\ket{00},\ket{10},\ket{11})\}$ of two qubits.}
\label{fig:shiftdown}
  \end{center}
\end{figure}
Using $\sigma_x$ and two CNOT gates one can construct shift down operation in the $V_3$ subspace as it is illustrated in Fig.~\ref{fig:shiftdown}.
In the retrieving part we can perform the controlled shift down gate (see Eq. (\ref{def:cshiftdown})) by simply adding a control qubit to those three gates (see Fig. \ref{fig:2na1toffoli}). The resulting quantum circuit for $2\rightarrow1$ PSR of phase gates contains  $2$ Toffoli gates, $2$ CNOT gates, and $3$ fixed one qubit gates. The success or failure of the retrieval is determined by the outcomes of the measurement of second and third qubit in the computational basis.
Outcome $01$ never appears, $11$ corresponds to failure and $00,10$ signalize successful retrieval of the phase gate. One can verify by a direct calculation that the success probability is $2/3$ and the related post-measurement state is $U_{\varphi}\ket{\xi}$.

\begin{figure}
  \begin{center}
    \includegraphics[width=8.5cm]{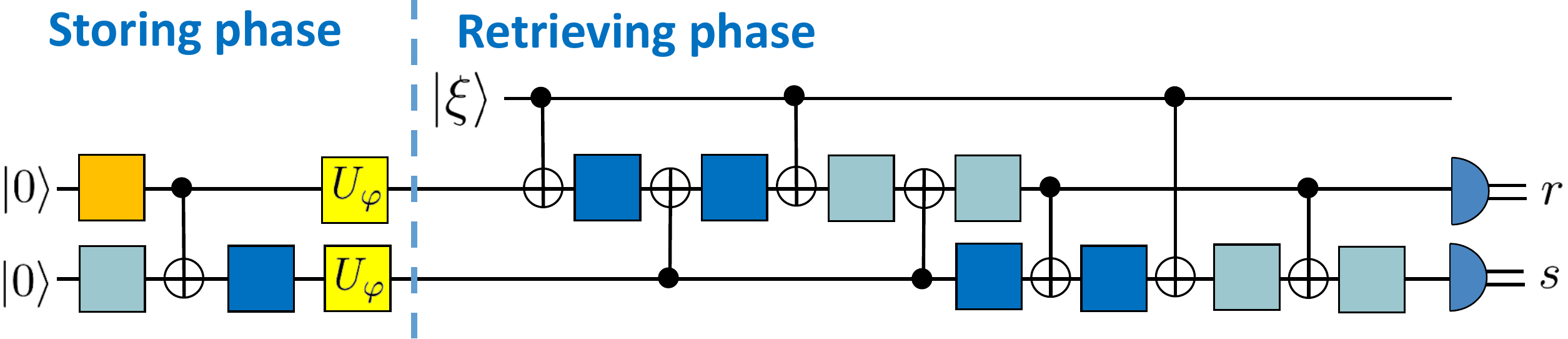}
\caption{Optimal $2\rightarrow 1$ PSR of phase gates using $8$ CNOT gates and $11$ fixed one-qubit gates. The color coding of one-qubit gates is the same as in Fig. \ref{fig:2na1toffoli}.}
\label{fig:2na1cnot}
  \end{center}
\end{figure}

Finally, the two Toffoli gates can be decomposed into elementary gates. Exact implementation of each Toffoli gate requires $6$ CNOT gates \cite{shendeToffcost}. However, we can be more efficient, because we have two Toffoli gates next to each other. We can employ a $3$-CNOT circuit (see \cite{barenco}, page $16$) that differs from Toffoli gate only by a phase of one state ($\ket{100}\mapsto -\ket{100}$). Luckily, this unwanted additional phase can be in our case cancelled (by suitable choice of the first and the second control qubit when using \cite{barenco}, page $16$) as we had two Toffoli gates next to each other. In this way $6$ CNOT gates can be saved. The resulting quantum circuit is depicted on Fig.~\ref{fig:2na1cnot}. One can verify by a direct calculation that the unitary transformation performed by the two Toffoli gates (from Fig.~\ref{fig:2na1toffoli}) is exactly reproduced by the last $6$ CNOT gates surrounded by $8$ one qubit gates in Fig.~\ref{fig:2na1cnot}. 
We conclude that we designed a quantum circuit containing $8$ CNOT gates and $11$ fixed one-qubit gates, which performs optimal $2\rightarrow 1$ PSR of phase gates.

\section{$(2^k-1) \rightarrow 1$ Realization: Efficient circuit}
\label{sec:2k-1PSRpg}
Designing a quantum circuit build from elementary gates and achieving optimal $N \to 1$ PSR of phase gates for arbitrary $N$ seems to be a challenging task. Already for $2\rightarrow1$ PSR of phase gates as we saw in the previous section it is not easy to find circuit containing low number of CNOT gates. For this reason it seems rather surprising if we manage to find the whole family of circuits with lowest possible complexity optimally realizing the task.

\begin{figure}
  \begin{center}
    \includegraphics[width=8.5cm]{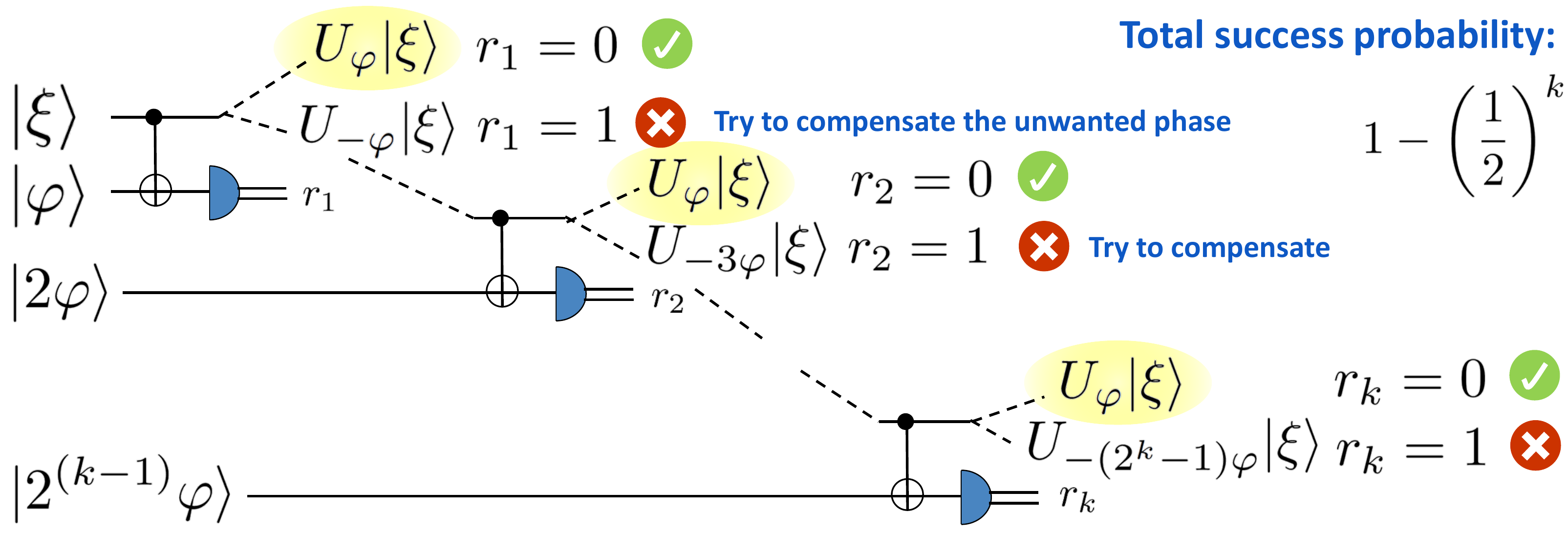}
\caption{Iterative use of programmable phase gate as proposed by Vidal, Masanes and Cirac in \cite{VMC1}.}
\label{fig:vmcscheme}
  \end{center}
\end{figure}

In section \ref{sec:1-1PSRpg} we used programmable phase gate by Vidal, Masanes and Cirac \cite{VMC1} to construct $1\rightarrow 1$ PSR of phase gates. From our perspective the main result of Vidal, Masanes and Cirac  in their paper \cite{VMC1} was actually to show that with suitable program states they can construct probabilistic phase gate, whose failure probability is exponentially small with respect to the number of qubits used as a program state. They proposed an iterative procedure, where the first step is a CNOT gate between the gate's input $\ket{\xi}$ and a program state $\ket{\varphi}\equiv\ket{\psi_\varphi}=(\ket{0}+e^{i \varphi}\ket{1})/\sqrt{2}$ (see Eq.(\ref{eq:CNOT11})). As we saw in section \ref{sec:1-1PSRpg} if measurement of the target qubit yielded bit $0$ (signalizing the success) then the gate's output was $U_{\varphi}\ket{\xi}$, otherwise they proposed to feed the output state $U_{-\varphi}\ket{\xi}$ again to their gate, but this time using the program state $\ket{2\varphi}$ (see Fig. \ref{fig:vmcscheme}). The gate would again succeed or fail with probability $1/2$, thus, after $k$ repetitions the success probability is $p_{\rm success}=1-1/2^k$.
In summary, this probability is achieved if the program register is prepared in state $\ket{\Psi^{(k)}_\varphi}\equiv \ket{\varphi}\otimes \ldots \otimes \ket{2^{k-1}\varphi}$.
Vidal, Masanes and Cirac \cite{VMC1} showed that their iterative scheme is optimal 
under the assumption that a program state is $\ket{\Psi^{(k)}_\varphi}$. However, the question whether this program state optimally encodes the phase gate $U_{\varphi}$ remained open.

In order to prepare the program state $\ket{\Psi^{(k)}_\varphi}$ one clearly needs
\begin{align}
1+2+\ldots+2^{k-1}=\sum_{m=0}^{k-1} 2^m = 2^k - 1
\end{align}
uses of the gate $U_\varphi$.

\begin{figure}
  \begin{center}
    \includegraphics[width=7cm]{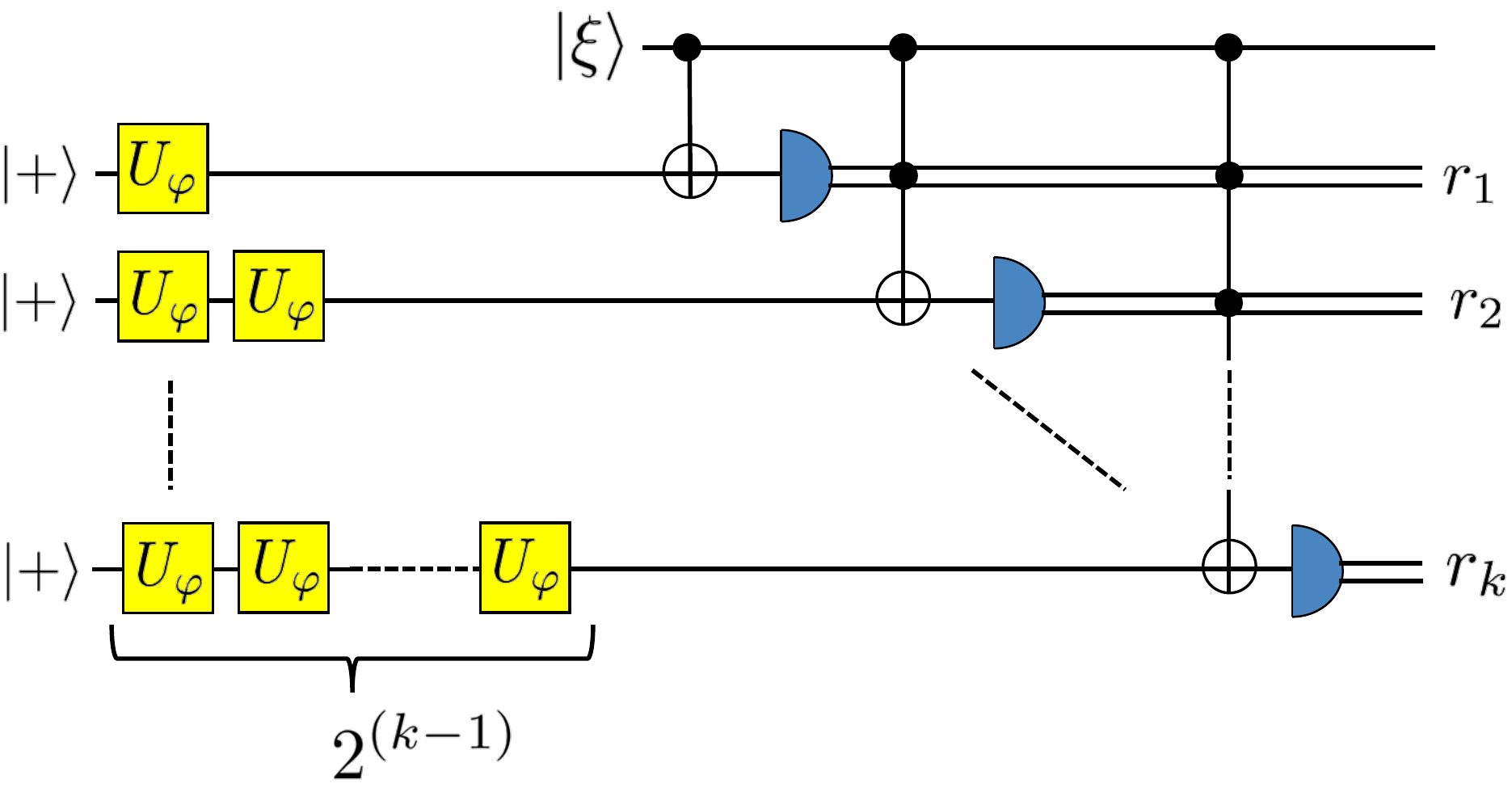}
\caption{Optimal $(2^k-1)\rightarrow 1$ PSR of qubit phase gates using 
$k$ CNOT gates and $k$ one-qubit measurements.}
\label{fig:2kna1}
  \end{center}
\end{figure}

Our theorem \ref{thm:mainclaim} implies that any procedure using the $U_\varphi$ gate $N=2^k-1$ times to probabilistically store and retrieve one use of $U_\varphi$ can succeed with probability at most $N/(N+1)=1-1/2^k$. This means that preparation of $k$ qubits in the state $\ket{+}^{\otimes k}$, production of state 
$\ket{\Psi^{(k)}_\varphi}$ by $2^k -1$ fold application of $U_\varphi$ gate and iterative application of programmable phase gate by Vidal, Masanes and Cirac \cite{VMC1} constitutes (see Fig. \ref{fig:2kna1}) a realization scheme for an optimal $(2^k-1) \rightarrow 1$ PSR of phase gates.
The clear advantage of the realization scheme described above is that it requires only $k$ CNOT gates and $k$ one-qubit measurements, while having exponentially small failure probability $1/2^k$ in the number of qubits $k$, which are used for the storage.

Can we do better? The symmetry of the problem is reflected by the commutation relations 
of the unitary group $U(1)$ as we detailed in the proof of Theorem 1.
Since each irrep of $U(1)$ acts in one-dimensional (complex) subspace, 
the effective system used to store and retrieve $U_\varphi$ is at least $D=N+1$
dimensional, if the performance should not be degraded and the phase gate is used $N$ times. It follows from the realization
proposed in Section \ref{sec:N-1PSRpg} that it is optimal with respect to
dimension needed. The $U(1)$ symmetry discussed above 
is shared with the problem 
studied by Vidal, Masanes and Cirac \cite{VMC1}.
Thus, the same dimension bound holds and requires the minimal dimension to be $2^k$, because in this case
$N=2^k-1$. Consequently, the whole Hilbert space of the system of $k$
qubits is needed to accomplish the task optimally. That is, under the assumption
of single-qubit measurements the qubit system in the state $\ket{\xi}$ and each of the $k$ qubits must be part of some interaction
in the retrieval phase, 
i.e. the process of retrieval consists of at least $k$ two-qubit gates.
In conclusion, any realization cannot use less
than $k$ CNOT gates, 
the described realization (Fig. \ref{fig:2kna1}) is indeed optimal also in the number of CNOT gates.
Moreover, our analysis from Section \ref{sec:optim-perf-learn}
answers positively the open question Vidal, Masanes and Cirac \cite{VMC1} had about the optimality of their covariant program state.

\tcb{
Finally, let us discuss the relation of the above scheme to the implementation proposed in Section \ref{sec:N-1PSRpg}. In principle, the intermediate measurements in Figure \ref{fig:2kna1} can be deferred to the last step at the expense of making the classical control quantum. In such case the $k$-th CNOT will become $k$-times controlled NOT gate. This possibility of rewriting the scheme (iterative procedure) as a fully quantum operation with measurement only at the end was noticed already by Vidal, Masanes and Cirac, but it is much less favourable for implementation by current quantum computer architectures.
It is worth noting that this fully quantum operation as a net result performs controlled shift down gate on a $2^k$ dimensional Hilbert space of $k$ qubits. Thus, scheme from Figure \ref{fig:2kna1} can be seen as a special case of the scheme from Section \ref{sec:N-1PSRpg}, where the realization can be split into iterative steps by making intermediate measurements.}

\section{Summary}
We addressed the question of efficient and optimal probabilistic 
storing and retrieving (quantum learning) of qubit phase gates $U_\varphi$. The learned information is stored in purely quantum way (as a suitable state of a quantum memory) and retrieved by means of programmable quantum processors.
In this paper we derived the optimal success probability for retrieval of one use of gate $U_\varphi$, if it was applied $N$ times in the storing phase, i.e. $N\to 1$ probabilistic storage and retrieval of qubit phase gates.
In comparison with the storage and retrieval of arbitrary qubit
gate \cite{ppsr1} the gain in the success probability decreases with
the number of uses. In particular, for storing and retrieval of arbitrary qubit unitary gate the optimal success probability equals $N/(N+3)$, whereas it equals $N/(N+1)$ if we restrict to phase gates only.

Further we investigated the question of efficient implementation of
the optimal storing and retrieving protocol. For the general $N\rightarrow 1$
case we designed a simple circuit realization exploiting a single controlled shift down gate, which is a generalization of CNOT gate to the case of $d$ dimensional (qudit) target system. In our case this qudit is $(N+1)$-dimensional and represents the effective space needed for the storage of the $N$ fold action of $U_\varphi$. For the case of $N=2$ we analyzed explicitly decomposition of all steps of the optimal protocol into elementary gates. We tried to minimize the number of CNOT gates and we found a $3$-qubit quantum circuit containing $8$ CNOTs that implements the optimal $2\to 1$ probabilistic storing and retrieving of qubit phase gates.

We argued that the programmable processor by Vidal, Masanes and Cirac \cite{VMC1} for programming of qubit unitary phase gates can be used to implement optimal $(2^k-1)\to 1$ probabilistic storage and retrieval of phase gates (see Fig. \ref{fig:2kna1}). Moreover, it follows from Theorem \ref{thm:mainclaim} and its proof
that the number of CNOT gates $k$ it uses is minimal and the considered program state is optimal. Hence, the open question left in \cite{VMC1} is answered positively. Let us also note that the storage performed as in Fig. \ref{sec:2k-1PSRpg} the phase gates are not employed
in parallel.
This adaptivity together with a special number $N=2^k-1$ of uses of the phase gate allows the size of the program system to coincidence with its theoretical minimum.





\acknowledgements{This work was supported by projects APVV-18-0518 (OPTIQUTE), VEGA 2/0161/19 (HOQIT) and QuantERA project HIPHOP. MS  was supported by The  Ministry  of  Education,  Youth  and  Sports  of the   Czech   Republic   from   the   National   Programme of   Sustainability   (NPU   II);   project   IT4 Innovations excellence in science - LQ1602. MZ acknowledges the support of project MUNI/G/1211/2017(GRUPIK). This publication was made possible through the support of the ID number 61466 grant from the John Templeton Foundation, as part of the “The Quantum Information Structure of Spacetime (QISS)” Project (qiss.fr). The opinions expressed in this publication are those of the author(s) and do not necessarily reflect the views of the John Templeton Foundation.}

\appendix

\section{Proof of Eq. (\ref{eq:reformperfectlearn})}
\label{sec:proof-oflemma1}
For any $J$ we define operator
$R_s^{(J)}:= I_J \otimes I_J \otimes s^{(J)}$.
We will perform direct calculation to evaluate $\bra{\psi} R_s^{(J)} \ket{\psi}$.
Let us denote the basis vectors of the $U(1)$ one-dimensional irreps and the vectors related to the decompositions (\ref{eq:decompopartial}) as follows
\begin{align}
\nonumber
  &\ket{v_j}_A\otimes\ket{0}_C = \ket{w_{j}}\otimes\ket{j} \nonumber \\
  &\ket{v_{j+1}}_A\otimes\ket{1}_C = \ket{w_{j}}\otimes\ket{j+1},
  \end{align}
where $\ket{v_j}\in \hilb{H}_j$, $\ket{w_J}\in \hilb{H}_J$, $\ket{j}\in \hilb{H}_{m^{(j)}_{j}}$, $\ket{j+1}\in \hilb{H}_{m^{(j+1)}_{j}}$.
Similarly we have,
\begin{align}
\nonumber
  &\ket{v_j}_{A'}\otimes\ket{0}_D = \ket{w_{j}}\otimes\ket{j} \nonumber \\
  &\ket{v_{j+1}}_{A'}\otimes\ket{1}_D = \ket{w_{j}}\otimes\ket{j+1},
  \end{align}
where $\ket{v_j}\in \hilb{H}_j$, $\ket{w_K}\in \hilb{H}_K$, $\ket{j}\in \hilb{H}_{m^{(j)}_{j}}$, $\ket{j+1}\in \hilb{H}_{m^{(j+1)}_{j}}$.
In the above notation we have $\ket{\psi}_{AA'}=\bigoplus_{j=0}^{N} \sqrt{p_j} \ket{v_j}_A \otimes \ket{v_j}_{A'}$

For $J=K=-1$ and $J=K=N$ the multiplicity spaces $\hilb{H}_{m_{-1,-1}}$, $\hilb{H}_{m_{N,N}}$ are one-dimensional, thus $s^{(-1)}$ and $s^{(N)}$ are just numbers. Direct calculation gives
\begin{align}
\nonumber
&\bra{\psi} R_s^{(-1)}\ket{\psi}=p_0 s^{(-1)} \ket{1}\bra{1}_C \otimes \ket{1}\bra{1}_D \nonumber\\
&\bra{\psi} R_s^{(N)}\ket{\psi}=p_N s^{(N)} \ket{0}\bra{0}_C \otimes \ket{0}\bra{0}_D \;,
\end{align}
which are operators not proportional to $\KetBra{I}{I}_{CD}=(\ket{0}\ket{0}+\ket{1}\ket{1})(\bra{0}\bra{0}+\bra{1}\bra{1})$.
Thus, we conclude that perfect storing and retrieving condition (see Eq. (\ref{eq:perfectlearcondJJ})) requires $s^{(-1)}=s^{(N)}=0$.

For $J=K=0,\ldots, N-1$ $s^{(J)}$ is an operator in $4$ dimensional multiplicity space. Due to Eq. (\ref{eq:diagonalR}) $s^{(J)}$ has only four nonzero elements, which we mark in the following way
\begin{align}
s^{(J)}=\sum_{a,b\in\{J,J+1\}} s^{(J)}_{a,b} \ket{a}\ket{a}\bra{b}\bra{b},
\end{align}
where $\ket{a}\ket{a},\ket{b}\ket{b},\in \hilb{H}_{m_{JJ}}$ (see Eq.(\ref{eq:decompototal})) and we remind that $\Ket{I_{m^{(J)}_J}}=\ket{J}\ket{J}$, $\Ket{I_{m^{(J+1)}_J}}=\ket{J+1}\ket{J+1}$.
Direct calculation for $J=0,\ldots, N-1$ then gives
\begin{align}
\nonumber
\bra{\psi} R_s^{(J)}\ket{\psi}&=p_{J} s^{(J)}_{J,J}\ket{00}\bra{00} +p_{J+1} s^{(J)}_{J+1,J+1}\ket{11}\bra{11} \nonumber\\
                            &+\sqrt{p_{J}p_{J+1}} \left(s^{(J)}_{J,J+1} \ket{00}\bra{11} + s^{(J)}_{J+1,J} \ket{11}\bra{00}\right),
\end{align}
which is proportional to $\KetBra{I}{I}_{CD}$ if and only if $s^{(J)}_{j,j'}=\mu_J/\sqrt{p_j p_{j'}}$.
Here $\mu_J$ is some number, which must be non-negative due to positive-semidefiniteness of $R_S$.




\begin{thebibliography}{99}
\bibitem{shor}
P. Shor,
 SIAM J. Computing 26 (1997), 1484-1509.

\bibitem{nielsen1}
M. A. Nielsen and Isaac L. Chuang,
Phys. Rev. Lett. 79, 321 (1997)


\bibitem{dariano2005}
G. M. D'Ariano and P. Perinotti,
Efficient Universal Programmable Quantum Measurements,
Phys. Rev. Lett. 94, 090401 (2005)

\bibitem{processor}
M. Hillery, V. Bu\v zek,and M. Ziman,
Phys. Rev. A 65, 022301 (2002)

\bibitem{kubicki}
A. M. Kubicki, C. Palazuelos, D. Perez-Garcia,
Phys. Rev. Lett. 122, 080505 (2019)


\bibitem{multimeter}
Miloslav Du\v sek and Vladimir Bu\v zek
Phys. Rev. A 66, 022112 (2002) 

\bibitem{multimeter1}
J. Fiur\'a\v sek, M. Du\v sek, and R. Filip
Phys. Rev. Lett. 89, 190401 (2002) 

\bibitem{garcia2006}
D. Perez-Garcia,
Optimality of programmable quantum measurements
Phys. Rev. A 73, 052315 (2006).


\bibitem{portishizaka0}
S. Ishizaka and T. Hiroshima, Phys. Rev. Lett. 101, 240501 (2008)

\bibitem{portIshizaka1}
S. Ishizaka and T. Hiroshima, Phys. Rev. A 79, 042306 (2009)

\bibitem{portStrelchuk1}
M. Studziński, S. Strelchuk, M. Mozrzymas, and M. Horodecki,
Sci. Rep. Vol. 7, 10871 (2017)
\
\bibitem{portstrelchuk2}
M. Mozrzymas, M. Studziński, S. Strelchuk, and M. Horodecki,
New J. Phys., Vol. 20, (2018)

\bibitem{ppsr1}
M. Sedl\'ak, A. Bisio, M. Ziman, 
Phys. Rev. Lett. 122, 170502 (2019)

\bibitem{bisilearn}
A. Bisio, G. Chiribella, G. M. D'Ariano, S. Facchini, P. Perinotti,
Phys. Rev. A 81, 032324 (2010)


\bibitem{VMC1}
G. Vidal, L. Masanes, J.I. Cirac,
Phys. Rev. Lett. 88, 047905 (2002)


\bibitem{comblong}
 G. Chiribella, G. M. D'Ariano, P. Perinotti,
Phys. Rev. A {\bf 80}, 022339 (2009).

\bibitem{architecture}
G. Chiribella, G. M. D'Ariano, P. Perinotti,
Phys. Rev. Lett. {\bf 101}, 060401 (2008).

\bibitem{supermaps}
 G. Chiribella, G. M. D'Ariano, P. Perinotti,
Europhysics Letters { \bf 83}, 30004 (2008).



\bibitem{barenco}
A. Barenco, C. H. Bennett, R. Cleve, D. P. DiVincenzo, N. Margolus, P. Shor, T. Sleator, J. A. Smolin, and H. Weinfurter,
Phys. Rev. A 52, 3457 (1995)

\bibitem{shendeToffcost}
Vivek V. Shende, Igor L. Markov, 
Quant.Inf.Comp. 9(5-6):461-486 (2009)


\end{thebibliography}


\end{document}